# All That's Happening behind the Scenes: Putting the Spotlight on Volunteer Moderator Labor in Reddit

**Hanlin Li[1], Brent Hecht[1], Stevie Chancellor[2]**

[1] Northwestern University
[2] University of Minnesota Twin Cities
lihanlin@u.northwestern.edu, bhecht@northwestern.edu, steviec@umn.edu

**Abstract**

Online volunteers are an uncompensated yet valuable labor force for many social platforms. For example, volunteer content moderators perform a vast amount of labor to maintain online communities. However, as social platforms like Reddit favor revenue generation and user engagement, moderators are under-supported to manage the expansion of online communities. To preserve these online communities, developers and researchers of social platforms must account for and support as much of this labor as possible. In this paper, we quantitatively characterize the publicly visible and invisible actions taken by moderators on Reddit, using a unique dataset of private moderator logs for 126 subreddits and over 900 moderators. Our analysis of this dataset reveals the heterogeneity of moderation work across both communities and moderators. Moreover, we find that analyzing only visible work – the dominant way that moderation work has been studied thus far – drastically underestimates the amount of human moderation labor on a subreddit. We discuss the implications of our results on content moderation research and social platforms.

## Introduction

Online volunteers are crucial to the success of prominent commercial social platforms, such as Reddit, Twitch, and Facebook Groups. Beyond all the publicly visible labor they do generating content, volunteers also perform managerial tasks behind the scenes such as content moderation, fact-checking, and norm-setting. This work ensures the health and vibrancy of social platforms and is essential for maintaining online communities.

Despite volunteers' utmost importance to many social platforms, they are not always the group that platforms prioritize in design and development, especially for volunteer content moderators. Prominent news outlets have reported that social platforms powered by volunteer moderators such as Reddit and Facebook prioritize revenue-generating user engagement over meeting volunteer moderators' needs (Peck, 2019; Washington Post, 2020). Moderators often feel under-appreciated, under-supported, and under-compensated by the platforms that rely on their labor (Gilbert, 2020; Matias, 2016; Postigo, 2009). This tension between moderators and the platforms they support boils over into public disagreements and disputes, e.g. "blacking out" popular communities on Reddit by making their content private and class-action lawsuits against AOL (Centivany and Glushko, 2016; Matias, 2016; Postigo, 2009).

To properly support moderators and preserve the online communities they maintain, the design and development of social platforms must be rooted in a comprehensive understanding of this labor. Existing approaches to researching content moderation at a large scale focus primarily on moderator activities that leave public visible traces, i.e. removing content and communicating with communities publicly,[1] (Chandrasekharan et al., 2019; e.g. Fan and Zhang, 2020; Jhaver et al., 2019c). However, new research shows that additional work happens behind the scenes such as managing user behavior and maintaining community settings (Gilbert, 2020; Lo, 2018). Without accounting for moderator labor as a whole, developers and researchers of social platforms risk undervaluing and driving away these volunteers and potentially undermining their platforms.

In this paper, we seek to more completely quantify and characterize moderator behaviors on Reddit. Working with Reddit moderators directly, we collected private moderator logs, called mod logs, from over 900 moderators of 126 subreddits. Private mod logs capture many more moderator actions in addition to the publicly visible ones mentioned above. As such, our dataset allows us to study the work that has fallen through the cracks of prior work and to build a taxonomy of visible and invisible work in content moderation. This broader lens shows that content moderation work is heterogeneous across both the subreddits and across moderators who work on the same subreddit. Moreover, despite being the main area of inquiry for moderation research, comment removal does not paint the full picture of content moderation; it may account for as little as 2% of total human labor across subreddits.

---

[1] Distinguished comments and posts will appear with a moderator badge, indicating that they are coming from a moderator.

For research efforts on content moderation and online communities specifically, our study complicates prior assumptions about moderator behavior and highlights the limitations of analyzing only visible moderation work. Our work also highlights the richness and scale of the volunteer labor that has helped to enable research and development beyond Reddit by improving user-generated data, e.g. large language models and mental health research (Brown et al., 2020; Chancellor et al., 2018). We discuss potential ways for researchers, developers, and labor advocates to understand and support this hidden labor in computing more comprehensively.

## Related Work

### Human Labor in Computing Systems

Science and Technology Studies (STS) literature has long argued that understanding and supporting workers is the precursor for successful and sustainable computing systems (Grudin, 1988; Star and Strauss, 1999). These labor practices have been a core area of interest in social computing (Geiger and Halfaker, 2013).

For our interests, *background labor* has been identified as both vital to platform health and simultaneously challenging to study. Background labor is work that is essential for the daily operation and maintenance of systems but is often obscured or ignored by the same systems (Star and Strauss, 1999); content moderation work is a prominent type of background labor. To understand background labor, researchers commonly use qualitative methods, such as ethnography, interviews, and self-reported survey data. Suchman (1995) provided an example of how ethnography around "document coding"—document work completed in a law firm to support attorneys and often misperceived as unskilled, "mindless" labor—unveiled the skills and expertise required. More recently, Kriplean (2008) called for researchers interested in Wikipedia editor activities to study background labor on the site such as administrative actions and providing social support.

Using content moderation as a case study, we build on the valuable insights in prior qualitative work to characterize background work. Although ethnography provides rich details about work activities, this method does not easily scale to massive remote collaboration across thousands of people. Similarly, interviews and surveys cannot provide granular insights into action-level work activities and also have limitations with self-reporting biases (Ernala et al., 2020). , we collaborated with Reddit moderators to collect private mod logs to provide a more expansive picture of their work patterns and practices.

### Content Moderation in Social Media

Three branches of the growing literature on content moderation informed our work and guided our analysis.

**Invisibility** is a known characteristic of content moderation and complicates research of social platforms. Moderator actions are made visible to users through changes to the content of a site. For example, removing a comment will leave traces to non-moderators, because the comment's text will be replaced with "[removed]". Conversely, some moderator actions are not publicly visible on the site. There are limited data traces that signal the occurrence of these actions and corresponding work. For example, when moderators ban users from a subreddit, this action is only visible to the affected users and invisible to the broader community. Qualitative studies have highlighted that much moderation work is made invisible to non-moderators by platform affordances and design decisions (Gilbert, 2020; Lo, 2018). This leads to an important observation – what specific actions are not visible to public-facing people (like researchers) and how do they compare to visible work in volume? Given mod logs' expansive coverage of visible and invisible moderator activities, our study lays out a classification of granular visible and invisible actions in Reddit content moderation and quantifies their volume.

**Makeup** of human moderation work at a granular level is another area of inquiry that has been challenging to study due to the lack of quantitative data about moderators' specific activities. Because comment removal is publicly visible, most community members perceive human moderation work as primarily being content removal (Myers West, 2018). In contrast, qualitative studies have described the richness and heterogeneity of human moderator labor (Jhaver et al., 2019b; Seering et al., 2020, 2019). Our work further validates these assumptions and qualitative findings with an action-level analysis of moderator behaviors.

**Workload** is another key area of inquiry in moderation research (Chandrasekharan et al., 2019; Lin et al., 2017). The amount of work that moderators perform is hard to quantify due to the invisibility of their activities. To understand what types of moderators face heavier workloads and could benefit from tooling support, researchers have relied on self-reported information and proxy measures (Matias, 2019). In our study, we mapped our dataset of mod logs onto each subreddit's posts and comments and in doing so, we provided a quantification of moderators' workload per post and comment.

# Background and Methods

## Background

On Reddit, each community (called a subreddit) is run by a team of volunteer moderators. Reddit moderators can take many kinds of actions on a subreddit, including approving and removing comments and posts, modifying the visual style of a community, and banning users, among others. All actions that a moderator takes using the built-in moderator functions on Reddit are recorded through moderator logs, or "mod logs". Reddit mod logs are a private record of moderation actions. Figure 1 shows a Reddit-provided example of their format. These logs are only accessible to a subreddit's moderators through the Reddit user interface or the Reddit API. Mod logs are not editable and are updated in real-time as actions occur.

## Data Collection

We collected mod logs from two sets of subreddits: 1) a subset of subreddits affiliated with u/publicmodlogs and 2) subreddits recruited by our research team. u/publicmodlogs is a Reddit bot that publishes all mod logs of subreddits to which it is installed. We included 84 subreddits from this list that were active at the time of our data collection, i.e. having at least one user post per day and one user comment per day when we gathered our data between June 2020 and January 2021). Because these 84 subreddits often cover niche topics (cryptocurrency, Not Safe for Work [NSFW] communities, and those with strong anti-censorship views), they may provide limited information about Reddit moderator behaviors more generally. In particular, this dataset does not include any large subreddits. To address this limitation, we directly recruited subreddits to contribute mod logs. We randomly sampled 400 subreddits using Reddit's `r/random` function and contacted their moderator teams through moderator mail, a private message channel that reaches all moderators on a subreddit. 42 subreddits' moderator teams shared their mod logs to support our research. This set of subreddits included three large communities that have over one million subscribers. We worked with moderators to determine what types of information should be anonymized or omitted during our data collection. We make available our data collection script for those interested in advancing the study of mod logs.[2] This part was reviewed by our Institutional Review Board for human subject research.

Given the sensitive nature of mod logs and the subsequent challenges in collecting this data, it was not realistic to capture a perfectly representative sample of moderators. Instead, we sought to develop a dataset that could catalyze progress towards unveiling and characterizing moderation work that may have been overlooked by researchers and developers previously. We note that in another study conducted using the same dataset, we compared the active moderators in our sample with the whole active moderator population using several publicly available activity metrics such as number of distinguished comments and account age. Although K-S tests show the distributions of these metrics differ between our sample and the population, means and medians are close. And the minimum and maximum values in our sample suggest that it also provides reasonable coverage in values (see Li et al., 2022 for more details). Put simply, our sample is an imperfect but somewhat representative sample of the moderator population.

---
[2] https://github.com/hanlinl/modresearch

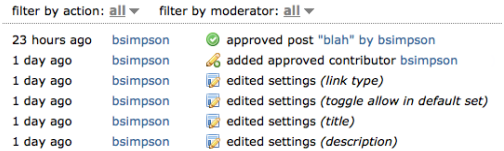

Figure 1: Reddit announced the mod logs feature in 2012 with this screenshot (https://www.reddit.com/r/mod-news/comments/nkj5s/moderators_moderation_log)

|  | Subscriber count in thousands | Daily average post count | Daily average comment count | Data collection span in days |
|---|---|---|---|---|
| Mean | 350+ | 70+ | 700+ | 142 |
| Max | 15,000+ | 2000+ | 20,000+ | 624 |
| 75% | 200+ | 40+ | 500+ | 169 |
| Median | 50+ | 15+ | 100+ | 167 |
| 25% | 20+ | 5+ | 20+ | 88 |
| Min | 5+ | 1 | 1 | 12 |

Table 1: An overview of our 126 subreddits' subscriber count, activity metrics, and data collection span.

## Dataset Overview

Our final dataset of mod logs includes over three million actions from 126 subreddits and over 900 moderator accounts (including both human and bot accounts). The dataset captured 64 types of moderation actions that go beyond approving and removing content actions and included editing subreddit Wikis or rules, adding flairs to posts, banning users, etc. To avoid confusion, we use the term "moderation" or the verb "moderated" to indicate that one of these 64 action types has been taken on posts or comments. We use the term "removed" to refer to posts or comments being removed by moderators.

Table 1 provides descriptive statistics about subreddits' subscriber count, daily post and comment counts, and timeframe. To protect moderators' and subreddits' anonymity, we reported all subreddits with an anonymized identifier. This is a combination of a subreddit's rank in subscriber count in our dataset and the category of its topical interest (out of news, gaming, politics, NSFW, and others). For example, r/1_humor is the largest subreddit in our dataset and focuses on humor-related content. The mean number of actions per day across the 126 subreddits is 25,812.

## Accounting for Invisible Work - A Taxonomy

We began our analysis by determining what work is visible for non-moderators. The outcome of content moderation may be discovered through API services like the Reddit API and Reddit's interface directly such as comment content being replaced with "[removed]". However, many moderation actions are not easily detected by users (if at all), and for those that can be detected, the impact of those changes may fade in a user's memory. For example, changing a subreddit's visual styles would be noticed only by users who recall that there was a previous version of the design; newcomers to a community would not "notice" this change at all. Such changes are not publicly logged anywhere except for the visual style itself, and it is likely that they would be "forgotten".

To help distinguish these levels of visibility, we create a taxonomy of visible and invisible labor. We draw on prior

| Actions (their visual representation on Reddit's user interface when applicable) | Daily occurrences by humans | Daily occurrences by bots |
|---|---|---|
| **3 - Invisible** | | |
| massive investigative efforts, or simply impossible to know | | |
| Approve content: | | |
|    approve comment | 1,017 | 43 |
|    approve post | 1,159 | 175 |
|    ignore report | 334 | 6 |
| Manage users: | | |
|    Ban/unban user (A private message titled "You have been banned from …" ) | 220 | 20 |
|    Mute/unmute user (A private message titled "You have been temporarily muted from …") | 24 | 23 |
|    add contributor | 5 | 8 |
| **2 – Potentially visible** | | |
| Some investigative effort, e.g. accessing the Reddit API or the Pushshift Reddit API to retrieve all removed posts and querying subreddit wiki pages periodically. | | |
| Remove posts: | | |
|    Remove post*, spam post* ( 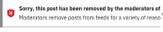 ) | 821 | 2,714 |
| Edit flairs/labels: | | |
|    Edit flair ( 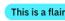 ) | 796 | 271 |
|    Mark nsfw, original, and spoiler (  ) | 41 | 48 |
| Change settings: | | |
|    Wiki revise | 95 | 481 |
|    Set how comments are sorted by default in a thread | 4 | 0 |
|    Edit rule | 3 | 0 |
| **1 – Easily visible** | | |
| No investigative efforts needed because direct cues are provided through UI | | |
| Remove comments: | | |
|    Remove comment, spam comment ( 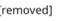 ) | 1,477 | 4508 |
| Engagement with communities: | | |
|    Distinguish | 450 | 3,615 |
|    Sticky | 235 | 5256 |
|    Lock | 137 | 498 |

*A post removed or labeled as spam by moderator accounts will still be available for direct visits via URL but its content will be filled with "[removed]" (whereas a post removed by the author themselves will appear as "[deleted]"). The post will also disappear from the subreddit's front page.

Table 1: A taxonomy of invisible and visible moderation actions by human moderators and bots
(some rare actions are omitted for space reason)

work by social computing scholars in social translucence (Erickson and Kellogg, 2000) and visibility as it applies to organizational systems. Specifically, Treem and Leonardi (2013) defined visibility as "the amount of effort people must expend to locate information". Their definition of visibility is relevant to content moderation because, similarly, technology design affects how visible moderators' work is to others who are not immediately privy to moderation actions. In such systems, accessible information that requires substantial efforts to retain is functionally invisible because people will be unlikely to look for it.

Following this approach, two authors of this paper consulted a content moderation researcher to classify each of the 64 moderation actions in terms of its visibility to non-moderators. Both authors involved are Reddit users and are very familiar with the overall UI of Reddit and APIs. One of the authors is a moderator of a medium-large community on Reddit. The content moderation researcher we consulted is also a moderator on a medium-large community on Reddit and has also worked closely with Reddit moderators.

Table 2 shows our taxonomy of invisible and visible moderation work on Reddit. Following Treem and Leonardi (2013)'s definition of visibility, the two authors and the researcher collectively mapped all moderation actions onto a three-point Likert scale. The scale corresponds to the amount of effort required for regular Reddit users as well as designers and researchers to know that a moderator action happened. An action is considered invisible (or rated a 3) if it is almost impossible for non-moderators to find any trace or the amount of effort required to get this information is impractical. For example, a user may be able to determine if a post was *approved* if it had appeared as "[removed]" before and they also remembered it. However, many posts are removed by `u/AutoModerator` *immediately* after their submission on the Reddit UI and API, making the task of tracking approvals functionally impossible. As such, we considered the "approve post" action to be invisible. In contrast, an action is rated as visible, or 1, if there are direct affordances in the Reddit UI and API that make the action obvious to all, like distinguishing comments or locking threads. Between invisible and visible actions, there exists a category of actions that are not immediately visible to users and researchers but may become visible with some investigative efforts, which we rate as 2. For example, post removals, although not shown on the front page of a subreddit, can be detected if users and researchers specifically search for removed posts via Reddit APIs or visit the post's URL.

Furthermore, under each level of visibility, the research team clustered actions based on what function they achieve, as also shown in Table 2. Under *invisible labor*, there are two thematic clusters, 1) approve content—actions that keep comments and posts up, and 2) manage users—actions that determine who could engage with a subreddit. Under *potentially visible labor*, there are 1) removing posts,[3] 2) edit flairs/labels—actions that assign posts categories but are not clearly labeled as moderator actions to users, and 3) change settings. Under *visible labor*, there are 1) remove comments and 2) engagement with communities.

Because automation is a key strategy for moderators to batch-moderate content, we separate bot actions from human moderators by drawing on prior approaches to bot detection (Jhaver et al., 2019c). We identified prominent bot accounts in our moderator lists such as `u/AutoModerator` and accounts whose names included words such as "bot" and "auto". In addition to this dictionary-based approach, we also identified extremely active accounts that performed more than 3000 moderation actions in one day, and manually inspected their profile pages to determine whether they were a bot. Many accounts identify themselves as bots in their posting history or profile page. For accounts about which we were uncertain, we contacted subreddits' moderator teams to ask if the account was a bot. In total, we classified 39 accounts as bots out of a total of 967 moderator accounts. Bots accounted for the majority (73%) of the 25,812 daily actions captured in our dataset.

**Visible and Invisible Labor**

Next, we move to examine the volume of invisible labor; in doing so we test prominent assumptions about moderator labor with this dataset. In this and the following sections, we use the format of stating prominent *unknowns or assumptions* from prior work and using our dataset to provide new insights or analyze whether the assumption holds.

**Who Does Invisible and Visible Work?**

**Unknown:** Qualitative evidence from interview studies has suggested that much of human moderators' work is invisible (Dosono and Semaan, 2019; Gilbert, 2020). However, these findings are based on interviews with moderators from one or a few subreddits and it is unclear whether these findings apply to a large, diverse set of subreddits. Prior work has also found visible traces of bots such as removed comments and removal explanations in comment threads (Jhaver et al., 2019b, 2019c). However, it remains unclear if bots are used for less visible types of work.

**Result:** Across subreddits, the share of invisible work for human moderators ranges from 9% to 94% with a median of 43%. Put another way, for half of the subreddits in our dataset, invisible work accounts for no less than 43% of human moderator labor. This quantitative evidence, therefore, supports prior qualitative findings on the invisibility of human labor in content moderation on a much larger scale (Gilbert, 2020; Lo, 2018). Conversely, the share of visible work varies from 2% to 68% with a median of 23%. For half of the subreddits in our dataset, visible work accounts for less than a quarter of all human labor. These results suggest human moderators indeed perform a significant amount of invisible work in addition to visible work.

---

[3] Removing posts is *invisible* because removed posts will disappear from a subreddit's front page, whereas removed comments will remain in comment threads with its content replaced with "[removed]".

With respect to the visibility of bots' work, we find that across subreddits, bots indeed perform visible work predominantly and are rarely used for invisible work. On average, 56% of bots' actions are visible (Median=62%) and only 6% (Median=2%) are invisible. Bots' focus on visible moderation actions further highlights the need for comprehensive analysis of human labor – if researchers and developers only examine visible work occurring in online communities such as removal, they may inadvertently count bots' work as human labor and overlook most human actions.

## Work Makeup

To gain a more granular understanding of moderator labor, we further investigate the makeup of work by bots and human moderators, respectively.

### Uniform Makeup of Bots' Work

**Assumption**: *Bots are primarily used to remove comments and posts and engage with comment threads.* Prior work has found that bots predominantly perform two categories of tasks: 1) removing comments and posts and 2) engaging with comment threads through distinguishing and/or "stickying" selected comments (distinguished comments will appear along with a moderator badge and stickied comments will appear at the top of the comment thread) (Jhaver et al., 2019b, 2019c). However, it is unclear if bots perform any additional work.

**Result:** Our dataset confirms this assumption and suggests that bots are rarely used for other types of moderation work. Across the total 18,843 daily bot actions (73% of 25,812 actions per day), removing comments and posts account for 38% of actions, and engaging with comment threads is 56% of bot work. The vast majority (94%, 118/126) of subreddits use bots to remove comments or posts, and just over half (52%) of subreddits use bots to automate distinguish/sticky actions to engage with comment threads. While there exist individual incidents of bots automatically updating subreddit wiki pages, this is rare in our dataset.

Our dataset provides additional insights on what types of subreddits have bots working on both content removal and engagement with threads. Notably, the subreddits that use bots for both purposes—content removal and engagement with threads—have a higher median subscriber count (Median=150,000+) than the subreddits that only used bots to remove comments or posts (Median=80,000+). A Mann-Whitney U test indicates that the difference was statistically significant (U=3079.0, p<0.05). Put simply, larger subreddits are more likely to use bots to automate both content removal and distinguish/sticky actions than smaller subreddits. Taken together, our analysis suggests that bots' use in content removal and engagement with comment threads is especially common among large subreddits.

### Heterogeneous Makeup of Human Labor

Figure 2(a) plots the percentage of each thematic cluster (defined in the Accounting for Invisible Work – A Taxonomy Section) relative to a subreddit's entire human moderation work for twenty subreddits. For a comprehensive overview of all the subreddits' makeup of human labor, see the Appendix. The left ten subreddits in the Figure are the ten

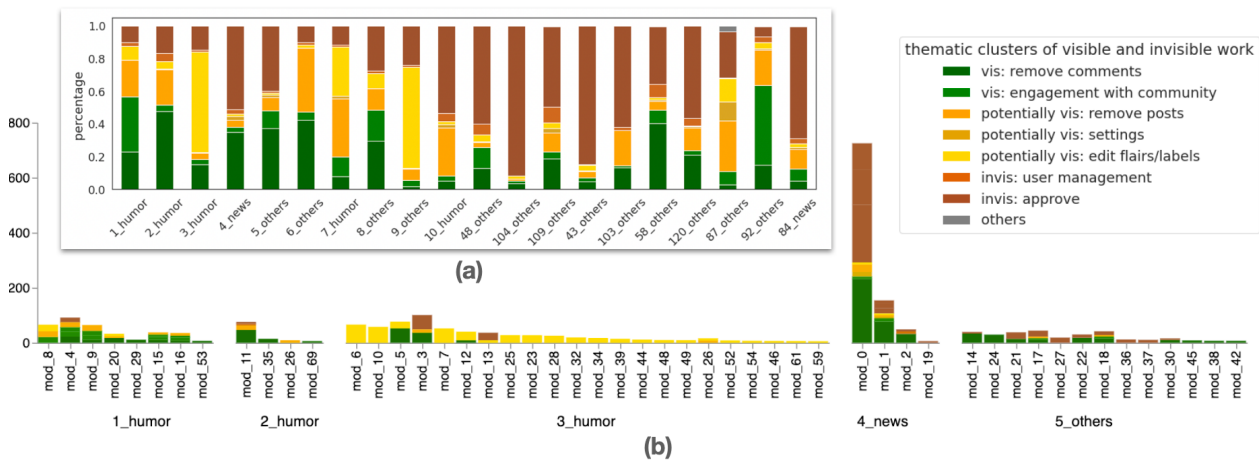

Figure 2: The distribution of human moderation work across the seven categories of moderation actions. (a)the distribution per subreddit. (b) the distribution per moderator

largest subreddits in our dataset by subscriber count (all with over half a million subscribers). We also included ten subreddits that have the highest volume of moderator activities *relative to subscriber count*, on the right. We focus on these ten subreddits because their moderator teams are the most active per subscriber and, therefore, offer distinctive insights into heavily moderated subreddits.

**Content Removal's Share of Human Labor**
**Assumption**: *Content moderation labor primarily consists of comment and post removal.* Prior work has shown that community members and the public perceive moderators' main responsibility as removing content (Myers West, 2018). Public discourse and media reports about content moderation also largely focus on removal and rarely touch on other aspects of this work (Facebook, 2021; Wired, 2014). Moreover, much prior work in supporting content moderation prioritizes removal, e.g. (Fan and Zhang, 2020; Jhaver et al., 2019a). In doing so, researchers and developers inadvertently reinforce the assumption that content removal is the primary component of moderation work.

**Result**: We find that comment and post removal accounts for 17% (`r/9_others`) to 74% (`r/2_humor`) of human labor among the ten largest subreddits in Figure 2(a) and 2% to 94% across all subreddits in our dataset. On more than half of all subreddits, comment and post removal accounts for less than 61% of overall human labor. These numbers complicate the assumption that comment and post removal is moderators' major responsibility because of how much it varies on subreddits. Furthermore, prior work that used removal-based traces of moderator labor such as removed comments is likely to underestimate "moderation volume" (Lin et al., 2017).

**Team-Level Heterogeneity**
**Assumption**: *Content moderation work is heterogeneous across subreddits.* While large social platforms such as Facebook have focused on developing generalizable tools to facilitate content moderation work, researchers have found that moderators of different communities have different values and approaches to their work (Chandrasekharan et al., 2018; Fiesler et al., 2018; Seering et al., 2019). For example, Fiesler et al. (2018) found that communities often express and enforce diverse rules, which imply different moderation practices behind the scenes, Whether this assumption holds has direct implications on what tasks researchers and developers of moderation tools focus on facilitating (Chandrasekharan et al., 2019).

**Result:** Returning to Figure 2(a), human moderators engage with diverse types of actions with different emphasis across subreddits. Specifically, each cluster of moderation actions makes up vastly different proportions of overall moderator labor across subreddits as seen in Figure 2(a). This can also be seen in the overview of the makeup of human labor in Appendix. For example, *approving content*, ranked at the top among nine subreddits' human moderators of the twenty subreddits in Figure 2(a), and 43 subreddits across our dataset (out of 126), accounts for as much as 92% in some subreddits' overall human labor, with a median of 34%. Similarly, *engagement with community* is ranked as the cluster accounted for the greatest percentage of actions on 16 subreddits in our dataset, with a range of 1-78% of human labor (median=6%). Moreover, unlike bots whose actions fall under *removing posts* and *comments* and *engagement with communities* primarily, human moderators cover all seven clusters of actions, regardless of to which subreddits they belong. These findings provide concrete evidence supporting prior work's finding on the diversity of moderator activities across subreddits, e.g. (Fiesler et al., 2018; Jhaver et al., 2019b; Seering et al., 2019).

**Individual-Level Heterogeneity**
**Assumption**: *Moderators of a given subreddit may perform different activities.* Prior interview studies with moderators have provided early evidence that moderators take on different roles (Jhaver et al., 2019b; Seering et al., 2020). For example, Seering et al. (2020) find a diversity of approaches in moderators' self-described philosophies. Therefore, it stands to reason that moderators may perform different types of actions in their day-to-day practices.

**Result**: Because larger subreddits tend to have larger moderator teams, we calculate the daily occurrences of the thematic clusters of actions per moderator for the five largest subreddits in our dataset and plot them in Figure 2(b). We find evidence supporting this assumption among these subreddits. On `r/1_humor`, while all moderators remove comments from this subreddit, some also take on other types of work, showing preferences towards *removing posts* (e.g. mod_8 and mod_9), some towards *approving content* (e.g. mod_4), and others towards *editing flairs/labels* (e.g. mod_8). On r/3_humor, human moderators consistently focus on *editing flairs/labels* and *approving content*; however, mod_3 and mod_5 also *remove comments*.

## Underlying Workload

Content moderation workload is an important metric that can inform future efforts to reduce human labor (Chandrasekharan et al., 2019). However, because moderation work leaves limited traces in public datasets, non-moderators have not yet comprehensively measured the volume of moderation work when studying community dynamics (Lin et al., 2017). Prior work has done so with proxy measures, like content removals in (Chancellor et al., 2016; Lin et al., 2017). In this section, we use mod logs to improve our understanding of the amount of work that bots and human moderators perform behind the scenes. We do so by

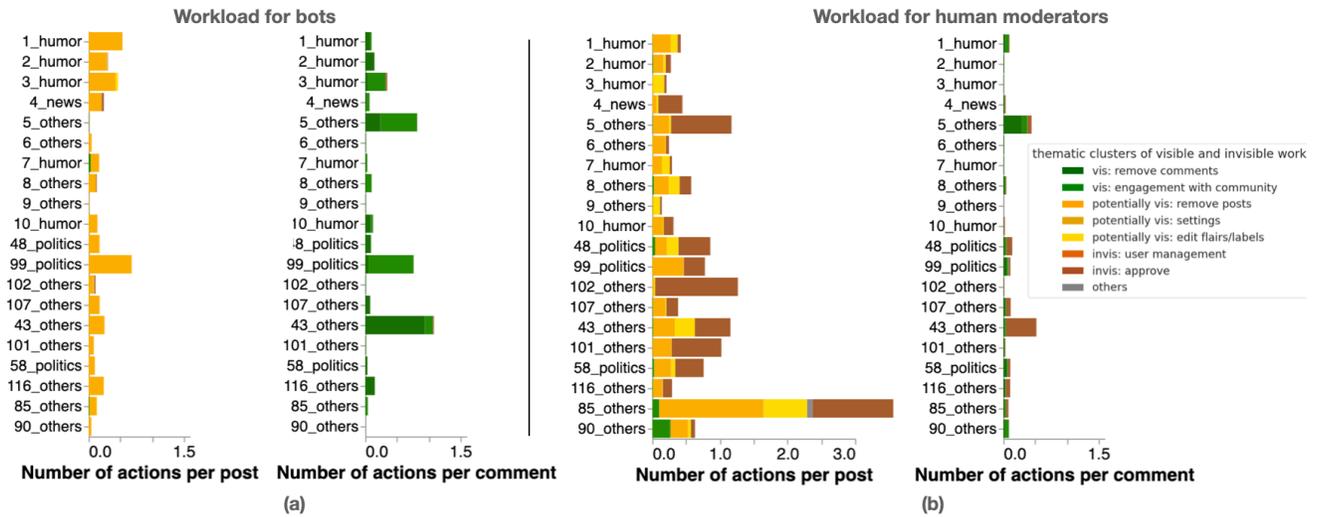

Figure 3: Workload for bots (a) and human moderators (b).

comparing our log data with all posts and comments returned by the Pushshift Reddit API. [4]

**Varied Workload for Bots**

**Unknown:** *To what extent do bots' workloads differ?* In previous work, some human moderators have reported that they emphasize reducing false negatives rather than false positives when using automation tools (i.e., using automation tools to catch as many posts and comments as possible) (Jhaver et al., 2019b). However, we have no current evidence about how this strategy plays out across subreddits.

**Result**: Overall, bots act on from 0% to 96% of posts and 0% to 45% of comments across our dataset. These wide ranges suggest that bots' role in shaping posts and comments varies greatly across subreddits.

Figure 3(a) plots the number of actions bots perform per post (left) and comment (right) among the same twenty subreddits from Figure 2. We find evidence showing moderators' extraordinary attempt to use bots to reduce false negatives on a few subreddits. On r/43_others, AutoModerator takes 0.90 remove comment actions per comment submitted. This means that most comments submitted to this subreddit were removed automatically and then manually screened for approval. In extreme cases, moderators may configure bots to manage all posts and comments on their communities.

Furthermore, we find that bots' workload varies between posts and comments; bots performed more actions per post than per comment with a few exceptions (such as r/5_others and r/43_others). Of the 109 subreddits in our dataset that use bots to moderate both posts and comments, 97 of them have bots performing more actions per post than per comment. We speculate that posts are more prominent than comments on Reddit's UI and, therefore, of higher stakes when moderators configure bots (Jhaver et al., 2019b).

**Varied Workload for Human Moderators**

**Unknown:** *To what extent do human moderator teams' workloads differ?* Like bots, the workload of human moderator teams' is hard to measure due to a lack of visibility into their actions. This hinders researchers' and practitioners' ability to concretely quantify the amount of human labor involved in supporting online communities.

**Result**: Figure 3(b) plots the number of actions human moderators take per post and comment, respectively, for the same twenty subreddits. Of the ten largest subreddits, r/5_others's human moderation workload is the heaviest, with each comment corresponding to 0.5 human actions and each post corresponding to 1.2 human actions. Across subreddits, human moderators perform, on average, 0.5 actions per post and 0.06 actions per comment. Human moderators have a material influence on posts and less influence on comments. This may be because actions on comments are more limited or that subreddits have more stringent rules for posts than for comments.

Like bots, human moderators' workload varies between posts and comments; humans focus their efforts on posts over comments relatively. On 120 subreddits, human moderators perform more actions per post than per comment. The disparity in workload in Figure 3(b) suggests that with the same amount of increase in posts or comments, different human moderator teams face different amounts of work.

---

[4] Notably, although it was previously unknown if the Pushshift API provides comments and posts that are quickly removed by bots such as AutoModerator, nearly all (93%) of the comments and posts removed by AutoModerator appeared in the Pushshift API, as either blank text or marked with "[removed]".

**Who Faces Greater Workload?**
**Assumption:** *Human moderators of larger subreddits face a greater workload per post and comment.* Knowing who performs more work is crucial for researchers and developers to prioritize and meet moderator needs. Prior work has argued that moderator work on larger subreddits is more important because of its potential to affect more Reddit users (Matias, 2019, 2016). Research efforts in understanding and supporting moderators also focus on larger subreddits (Chandrasekharan et al., 2019; Jhaver et al., 2019b). However, we do not know if human moderators of larger subreddits have heavier workloads per post or comment than those from small subreddits.

**Result**: We did not observe evidence supporting this hypothesis that larger subreddits have more human labor per post or comment. The workload of human moderators per post or comment is not associated with a subreddit's subscriber count (post: spearman's rho = -0.02, $p>0.05$; comment: spearman's rho = -0.08, $p>0.05$). Put another way, large subreddits' moderator teams do not have a heavier workload per post or comment than those of smaller subreddits, though they are likely to have a heavier workload absolutely (post: spearman's rho = 0.64, $p<0.001$; comment: spearman's rho = 0.67, $p<0.001$).

**Distribution of Workload**
**Unknown**: *How equally is moderation work distributed within a moderator team?* Prior interviews have offered insight into the different levels of involvement moderators have with their teams such as "the head mod" vs. "the janitor" (Seering et al., 2020). However, this prior work did not identify how equally they distribute work among themselves. Currently, in data-driven research on online communities, researchers treat all moderators equally by using moderator count in their analysis (Kiene and Hill, 2020; Matias, 2016). However, there may exist moderators who perform little work adding noise to such models.

**Result**: Returning to Figure 2(b), we find the distribution of moderation work among a subreddit's moderators is highly unequal. We further calculate the Gini index on moderator actions, a measure of inequality, for each of the 36 subreddits with ten or more human moderators. The Gini index values ranged from 0.47 to 0.90 (median = 0.74). Most prominently, in Figure 2(b), `r/4_news`'s most active moderator, mod_0, was responsible for 72% of all the moderation work on the subreddit. Taken together, a subreddit's moderation work likely concentrated on a few moderators, with the rest performing comparatively few actions.

# Discussion

## Implications on Content Moderation Research

Our findings on the invisibility and heterogeneity of content moderation complicate existing research approaches that rely on publicly available datasets to study moderators' labor activities. As human moderation actions are not always visible, methods that only assess publicly visible work, e.g. removing comments, will very likely leave out a significant portion of work that happens behind the scenes. They may also overlook differences in work makeup and workload across subreddits. Future work must contend with the invisibility and heterogeneity of moderation work if they wish to meaningfully engage with the full scope of moderator labor.

How could research build on our findings? For quantitative studies that characterize moderator engagement, researchers may take our taxonomy as a starting point to investigate invisible labor. Our results suggest that with more investigative efforts into collecting traces of moderation actions, researchers' ability to "see" these actions has the potential to improve accordingly. Additionally, our study suggests that researchers need to ensure the validity of moderator activity metrics across subreddits in their modeling of moderator behaviors given the heterogeneity of moderation labor. For example, metrics that signal strong moderator engagement on one subreddit (e.g. number of distinguished comments) may not work as well on another.

For qualitative work, our findings amplify complementary perspectives about the multiplicity of moderator work that moves beyond content removal (Gilbert, 2020; Seering et al., 2020). Future work may focus on moderator behaviors that have not yet been fully understood or supported by existing moderation tools. For example, moderators may benefit from tools that automatically approve certain content or users or edit flairs to ease some of their burdens. Our findings suggest that the landscape of content moderation is vastly diverse; no two subreddits are alike. Future work may further explore different ways of content moderation and construct archetypes of moderation strategies.

## Implications on Computing Research

Our work highlights the labor supporting the creation of large-scale Reddit datasets and the research communities that rely on Reddit for knowledge production. Reddit data is influential in computing and beyond (Baumgartner et al., 2020; Bevensee et al., 2020), supporting research on topics such as political extremism (Chandrasekharan et al., 2017; Farrell et al., 2019) and mental health (Chancellor et al., 2018; Choudhury and Kiciman, 2017), as well as contributing to powerful machine learning models such as GPT-3 (Brown et al., 2020). When researchers leverage large-scale, user-generated datasets for scientific research, the moderator labor involved in the production and curation of these datasets is poorly understood and documented, even though moderator labor directly influences posts and comments and potentially research outcomes. This documentation (and lack thereof) is especially worrisome when datasets are used for high-stake decision-making (Bandy and Vincent, 2021; Gebru et al., 2020; Geiger et al., 2020; Proferes et al., 2021). Gebru et al. (2020) have cautioned that without proper documentation of datasets, researchers may make false assumptions of representativeness or generalizability and research outcomes. Proferes et al. (2021) further pointed out that in the case of Reddit, two prominent contextual factors—community culture and demographics—may influence model generalizability. Indeed, at one extreme, `r/43_others'` moderators filtered the bulk of the comments submitted to this subreddit and, therefore, greatly influence what content is available on this subreddit. But `r/43_others` is no singular or novel outlier - as seen in Figure 3, there are several subreddits whose moderators frequently made decisions about what content to remove or approve and thereby, affect their subreddits' content availability. These findings affirm moderators' role in shaping user-generated content and highlight the importance of accounting for content moderation in dataset documentation and research more generally.

### Supporting Labor in Social Platforms

Our findings also problematize existing approaches that examine only the visible part of background labor—work that is essential to systems' operation and maintenance but often overlooked by those involved (Star and Strauss, 1999). Currently, data about the invisible part of background labor is largely held behind closed doors of private companies and is difficult to access for researchers.

Our data collection and analysis point to some potential directions for researchers to resolve these tensions. First, researchers may collaborate with workers directly (like moderators) and deploy tools that collect their log data with strict privacy protection. In the crowdwork domain, tools have been developed to allow crowdworkers to see their hourly wage and simultaneously quantify invisible, unpaid labor for researchers (Hara et al., 2018; Toxtli et al., 2021). Future work may explore how this approach could benefit uncompensated digital labor, such as volunteer content moderation and peer production while helping moderators conduct their own "time studies" (Khovanskaya et al., 2019).

Second, our results on the sheer volume of volunteer labor necessary to maintain online communities further highlight the importance of recognizing and supporting volunteer labor. Reddit moderators have long needed better support for their work as well as protection against the risks associated with their role such as online harassment (Gilbert, 2020; Matias, 2019, 2016). One factor contributing to their lack of negotiation power in their relationship with platforms is the invisibility of their labor and an inability to quantify their contributions (Li et al., 2022). Designers of computing systems could consider improving the visibility of moderation work to correct these misperceptions and focus internal resources to support invisible work (Suchman, 1995). This could be done through interface changes or public reporting such as "this subreddit's moderator team has worked 18 hours for the community in the past week". However, we strongly caution against wholescale attempts at making all invisible moderation work visible given the risks of social surveillance and harassment by bad actors (Gilbert, 2020). Any attempt that seeks to increase the visibility of moderation work needs to contend with the importance of moderators' privacy, safety, and wellbeing.

### Limitation

Although mod logs provide expansive coverage of moderator activities, there still exists invisible moderation work that is not present in mod logs. Two prominent examples are responding to mod mails and deliberation within moderator groups. Prior qualitative work has noted both the importance of this work and the challenges in capturing it (Dosono and Semaan, 2019; Gilbert, 2020). Our study did not characterize such activities given mod logs' limitations. There are other opportunities to understand, characterize, and support these untraceable moderator activities—a fruitful area for future research. One may explore working with moderators even more closely by conducting diary studies to address this limitation.

## Conclusion

Using Reddit moderation logs, we complicate prior assumptions about content moderation work and highlight how moderator labor has been partially overlooked or misunderstood. Specifically, we expose the amount of invisible labor in moderation and uncover heterogeneous work makeup and varied underlying workload. Our study highlights the importance of accounting for obscured human labor in content moderation and computing research in general that relies on Reddit data.

## Appendix

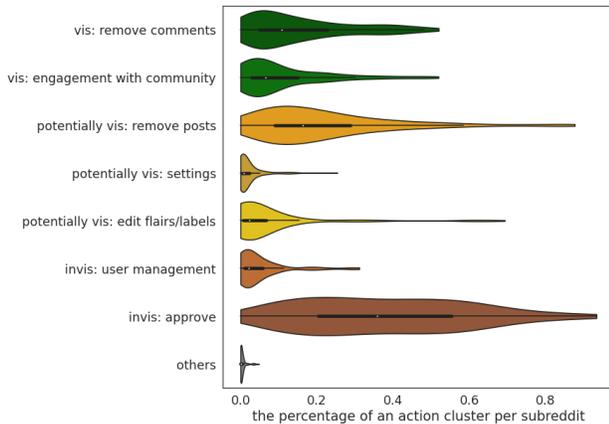

Figure 4: The distribution of the percentages of each action cluster across subreddits. These action clusters are defined in the Accounting Invisible Work – A Taxonomy section. Notably, the percentages of *approve content* vary widely across subreddits.

## Acknowledgments

The authors would like to thank Sarah Gilbert for providing feedback on an early draft of this paper as well as for helping us improve our taxonomy of visible and invisible moderation work. The authors would also like to thank all the moderators whose data made this study possible.